\documentclass[twocolumn,letter]{jpsj3}
\usepackage{txfonts}

\title{Evolution of the electronic state through the reduction annealing in electron-doped Pr$_{1.3-x}$La$_{0.7}$Ce$_x$CuO$_{4+\delta}$ ($x=0.10$) single crystals: \\Antiferromagnetism, Kondo effect and superconductivity}

\author{Tadashi Adachi$^1$\thanks{E-mail: t-adachi@sophia.ac.jp}\thanks{Present address: Department of Engineering and Applied Sciences, Faculty of Science and Technology, Sophia University, 7-1 Kioi-cho, Chiyoda-ku, Tokyo 102-8554, Japan}, Yosuke Mori$^1$, Akira Takahashi$^1$, Masatsune Kato$^1$, Terukazu Nishizaki$^2$\thanks{Present address: Department of Electrical Engineering and Information Technology, Kyushu Sangyo University, Fukuoka 813-8503, Japan}, Takahiko Sasaki$^2$, Norio Kobayashi$^2$, and Yoji Koike$^1$}
\inst{$^1$Department of Applied Physics, Graduate School of Engineering, Tohoku University, \\6-6-05 Aoba, Aramaki, Aoba-ku, Sendai 980-8579, Japan \\
$^2$Institute for Materials Research, Tohoku University, \\2-1-1 Katahira, Aoba-ku, Sendai 980-8577, Japan \\
} 

\abst{The evolution of the electronic state through the reduction annealing has been investigated in electron-doped Pr$_{1.3-x}$La$_{0.7}$Ce$_x$CuO$_{4+\delta}$ ($x=0.10$) single crystals with the so-called T' structure. 
From the ab plane and c axis electrical resistivity measurements in magnetic fields, it has been found that, through the reduction annealing, the strongly localized state of carriers accompanied by the antiferromagnetic (AF) pseudogap in the as-grown crystal changes to a metallic state bringing about the Kondo effect without AF pseudogap and to a superconducting state. 
These results are able to be understood in terms of a model based on the strong electron correlation. 
The complete removal of excess oxygen in the T'-cuprates is expected to result in the appearance of superconductivity in a wide range of the Ce concentration including the parent compound of $x=0$.}

\kword{T' structure, reduction annealing, superconductivity, Kondo effect, electron-doped cuprate}

\begin{document}
\maketitle


For the aim to elucidate the mechanism of superconductivity in the high-$T_{\rm c}$ cuprates, there have been intensive studies to establish the phase diagram. 
On the phase diagram of the electron-doped cuprate Nd$_{2-x}$Ce$_x$CuO$_4$ (NCCO) with the so-called T' structure, an antiferromagnetic (AF) order is formed in the underdoped regime of $x<0.14$.~\cite{luke}
It is well known that as-grown crystals with the T' structure tend to contain excess oxygen. 
The excess oxygen is understood to induce disorder of the electrostatic potential in the CuO$_2$ plane, leading to breaking of the Cooper pairs.~\cite{xu}
Therefore, the removal of the excess oxygen through the reduction annealing is indispensable to the appearance of superconductivity and also to the access of the intrinsic properties of the electron-doped cuprates.
On the other hand, recent neutron-scattering and x-ray diffraction studies have suggested that the reduction annealing brings about filling up Cu deficiencies in the as-grown crystals where the Cu deficiencies are regarded as causing the pair-breaking.~\cite{kang} 

Recently, Matsumoto {\it et al}. have reported a surprising result that properly reduced thin films of NCCO show superconductivity in a wide range of $x$ even at $x=0$.~\cite{matsumoto} 
Formerly, Brinkmann {\it et al}. have observed superconductivity even in the underdoped regime of $x \ge 0.04$ in Pr$_{2-x}$Ce$_x$CuO$_4$ (PCCO) through the improved reduction annealing.~\cite{brinkmann}
Obviously, these observations of superconductivity in the extremely underdoped regime are unable to be explained in terms of carrier doping into a parent compound of a Mott insulator as in the case of the hole-doped cuprates, being in sharp contrast to the common understanding of the well-known phase diagram of the electron-doped cuprates.

From the theoretical viewpoint, the superconductivity in the parent compounds of the electron-doped cuprates might be explained by the early local density energy band study in NCCO~\cite{massidda} or by the recent half-filled-band Hubbard model consisting of doublons and holons.~\cite{yokoyama}
These are based on the concept of a band metal without Mott-Hubbard gap due to the strong electron correlation.
On the other hand, Weber {\it et al}. have suggested from a calculation using the local-density approximation combined with the dynamical mean-field theory that, being contrary to the parent Mott insulators of the hole-doped T-cuprates, the parent compounds of the T'-cuprates are Slater insulators in which the insulating behavior is the result of the presence of magnetic order, although both T- and T'-cuprates are strongly correlated electron systems.~\cite{weber} 
That is, both metallic and superconducting (SC) states are expected to appear by eliminating the magnetic order in the parent compounds of the T'-cuprates. 
Accordingly, a critical question is whether or not the superconductivity in the parent compounds of the electron-doped cuprates appears based on the strong electron correlation.

Another fundamental question is about the origin of the AF order in the well-known phase diagram.
Since the reduction annealing causes the destruction of the AF order and the appearance of superconductivity, it is speculated that the AF order originates from the remaining excess oxygen in a crystal, although the details have not yet been clarified.
Onose {\it et al}. have found that the c axis electrical resistivity $\rho_{\rm c}$ of NCCO with $x<0.14$ exhibits a broad hump in a temperature range between 200 K and 400 K where the opening of pseudogap has been observed from their optical-conductivity measurements.~\cite{onose}
Since the pseudogap has been observed in crystals showing the AF order at low temperatures, they have pointed out that the AF fluctuation induces the pseudogap at the Fermi surface between ($\pi ,0$)/($0, \pi$) and ($\pi/2,\pi/2$) in $k$ space, resulting in the suppression of the increase in $\rho_{\rm c}$ with decreasing temperature, namely, the broad hump. 
However, the relationship between the broad hump of $\rho_{\rm c}$ and excess oxygen has not yet been clarified.
In the ab plane electrical resistivity $\rho_{\rm ab}$ measurements in magnetic fields for electron-doped non-superconducting thin films of NCCO and PCCO undergoing the reduction annealing, on the other hand, Sekitani {\it et al}. have found log $T$-dependence of $\rho_{\rm ab}$ and negative magnetoresistance at low temperatures, suggesting the occurrence of the Kondo effect due to free Cu spins adjacent to the slightly remaining excess oxygen.~\cite{sekitani}
However, the relationship between the excess oxygen, Kondo effect and pseudogap has been less understood in the electron-doped cuprates.

\begin{figure}
\includegraphics[width=1.0\linewidth]{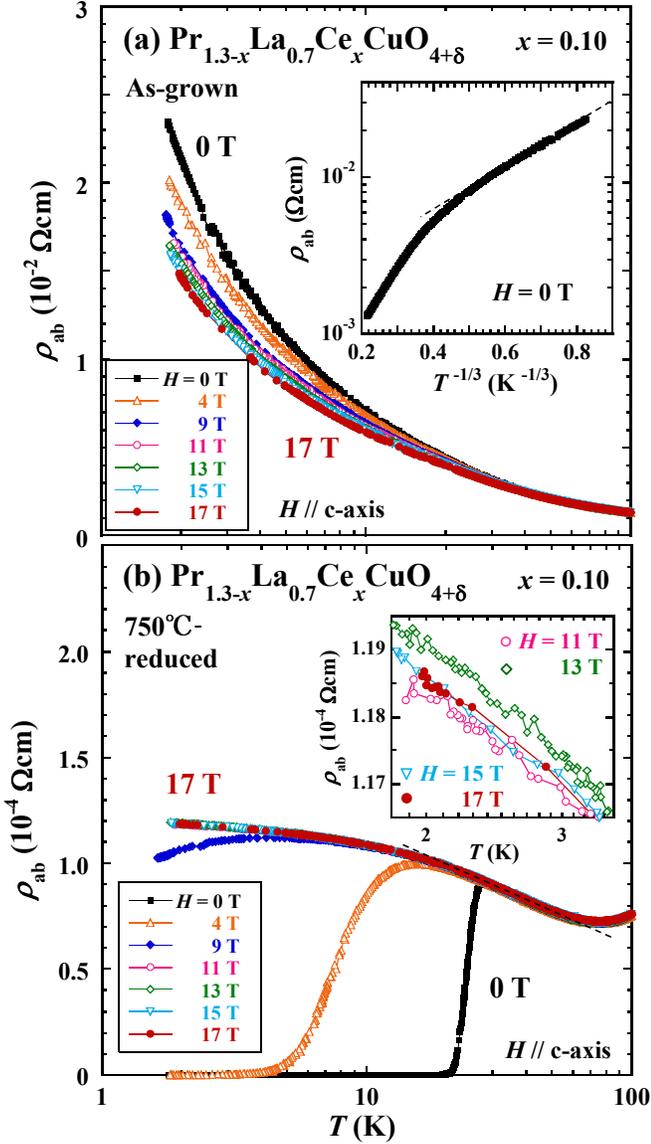}
\caption{(Color online) Temperature dependence of the ab plane electrical resistivity $\rho_{\rm ab}$ in magnetic fields parallel to the c axis on field cooling for (a) as-grown and (b) 750$^{\rm o}$C-reduced Pr$_{1.3-x}$La$_{0.7}$Ce$_x$CuO$_{4+\delta}$ with $x=0.10$. The inset of (a) is a plot of log $\rho_{\rm ab}$ vs. $T^{-1/3}$ in zero field.  The inset of (b) is a magnified plot of $\rho_{\rm ab}$ in magnetic fields above 11 T at low temperatures.}
\label{f1}
\end{figure}

In this Letter, we have investigated the change of the electronic and magnetic states in the CuO$_2$ plane through the reduction annealing from $\rho_{\rm ab}$ in magnetic fields and $\rho_{\rm c}$ in zero field using electron-doped single crystals of Pr$_{1.3-x}$La$_{0.7}$Ce$_x$CuO$_{4+\delta}$ (PLCCO) with $x=0.10$.
It has been found in the as-grown crystal that carriers are strongly localized at low temperatures and the AF pseudogap is open.\
For the moderately reduced crystal, on the other hand, $\rho_{\rm ab}$ has been found to exhibit the Kondo effect at low temperatures and the AF pseudogap is closed.
These results suggest that the AF order is induced by the excess oxygen in the electron-doped cuprates.
We propose a novel band structure based on the strong electron correlation well explaining the present results and discuss the evolution of the electronic and magnetic states in the CuO$_2$ plane through the reduction annealing.


Single crystals of PLCCO with $x=0.10$ were grown by the traveling-solvent floating-zone method in air.~\cite{koike,fujita,risdi} 
The quality of the crystals was checked by x-ray back-Laue photography and x-ray diffraction to be good.
The composition of as-grown crystals was analyzed by the inductively coupled plasma (ICP) analysis to be Pr$_{1.17}$La$_{0.73}$Ce$_{0.10}$Cu$_{1.00}$O$_{4+\delta}$ which is close to the nominal composition. 
As mentioned above, Kang {\it et al}. have suggested that the effect of the reduction annealing is to fill up the Cu deficiencies in the as-grown crystal and increase the conductivity.~\cite{kang}
However, our ICP result of the as-grown crystal proved the absence of Cu deficiencies within an error. 
Moreover, the removal of excess oxygen through the reduction annealing was confirmed by the weight change in Pr$_{1-x}$LaCe$_x$CuO$_4$.~\cite{fujita,risdi}
Scanning electron microscope (SEM) also revealed that, in the reduced SC crystal, neither Cu metals nor rare-earth oxides were observed. 
Therefore, it is doubtful that the present results are explained in terms of the Cu-deficiency model.

For the reduction annealing, we performed annealing in vacuum of $\sim10^{-6}$ torr for 24 hours at various temperatures.
In order to protect the surface of crystals during the annealing, as-grown crystals were covered with polycrystalline powders having the same composition. 
This is called protect annealing, which is an improved version of the technique used by Brinkmann {\it et al}.~\cite{brinkmann}
Although the value of $\delta$ was not determined exactly, the decrease in the c axis lattice constant was observed through the reduction annealing, suggesting the removal of excess oxygen.

The $\rho_{\rm ab}$ and $\rho_{\rm c}$ measurements were performed by the standard dc four-probe method. 
The $\rho_{\rm ab}$ was measured in magnetic fields parallel to the c axis up to 17 T on field cooling. 
The magnetic-susceptibility measurements were carried out in a magnetic field of 5 Oe parallel to the c axis.


In Fig. 1, $\rho_{\rm ab}$ vs. log $T$ in various magnetic fields parallel to the c axis is plotted for as-grown and 750$^{\rm o}$C-reduced crystals of PLCCO with $x=0.10$. 
In zero field, the as-grown crystal exhibits semiconducting temperature dependence and no trace of superconductivity.
For the crystal reduced at 650$^{\rm o}$C (not shown), on the other hand, $\rho_{\rm ab}$ is semiconducting at high temperatures, while a drop of $\rho_{\rm ab}$ is observed below $\sim 7$ K due to the SC transition. 
For the crystal reduced at 750$^{\rm o}$C, $\rho_{\rm ab}$ shows a metallic behavior at high temperatures and an upturn at low temperatures below $\sim$ 75 K, followed by the SC transition below 20 K.
In fact, Meissner diamagnetism due to bulk superconductivity was detected below $\sim 20$ K for this crystal.
Formerly, Sun {\it et al}. have reported the transport properties of PLCCO single crystals reduced through the annealing in Ar that a crystal with $x=0.13$ is SC while crystals with $x \le 0.10$ are non-SC.~\cite{sun}
The present successful observation of bulk superconductivity in PLCCO with $x=0.10$ suggests that the excess oxygen is effectively removed from the as-grown crystal through the protect annealing.

The $\rho_{\rm ab}$ of the as-grown crystal exhibits neither log $T$ dependence nor saturation at low temperatures. 
Instead, as shown in the inset of Fig. 1(a), log $\rho_{\rm ab}$ is proportional to $T^{-1/3}$ at low temperatures, suggesting that carriers are strongly localized and that two-dimensional variable-range-hopping conduction is realized. 
Moreover, $\rho_{\rm ab}$ below $\sim 40$ K is suppressed by the magnetic field, that is, negative magnetoresistance is observed.
For the 750$^{\rm o}$C-reduced crystal, on the other hand, $T_{\rm c}$ decreases by the magnetic field and the superconductivity disappears above 11 T.
It is found that $\rho_{\rm ab}$ exhibits log $T$ dependence in a wide temperature range between 20 K and 50 K and tends to be saturated at $\sim$ 2 K above 11 T.
Moreover, negative magnetoresistance is observed at low temperatures above 13 T, as shown in the inset of Fig. 1(b).
These are characteristic of the Kondo effect, suggesting that the strongly localized state of carriers changes to a metallic state with the Kondo effect through the reduction annealing in PLCCO with $x=0.10$.

Figure 2 shows the temperature dependence of $\rho_{\rm c}$ in zero field for PLCCO with $x=0.10$.
For as-grown and 650$^{\rm o}$C-reduced crystals, a hump is observed at $\sim$ 200 K, which is similar to that observed in NCCO with $x<0.14$~\cite{onose} and PLCCO with $x=0.01$ and 0.03~\cite{sun} and due to the opening of AF pseudogap.
For the 750$^{\rm o}$C-reduced crystal, on the contrary, a simple metallic behavior without hump is observed, which resembles the behavior of $\rho_{\rm c}$ in SC NCCO with $x=0.15$.~\cite{onose} 
Therefore, it is inferred that the AF pseudogap disappears in the SC crystal reduced at 750$^{\rm o}$C.

\begin{figure}
\includegraphics[width=1.0\linewidth]{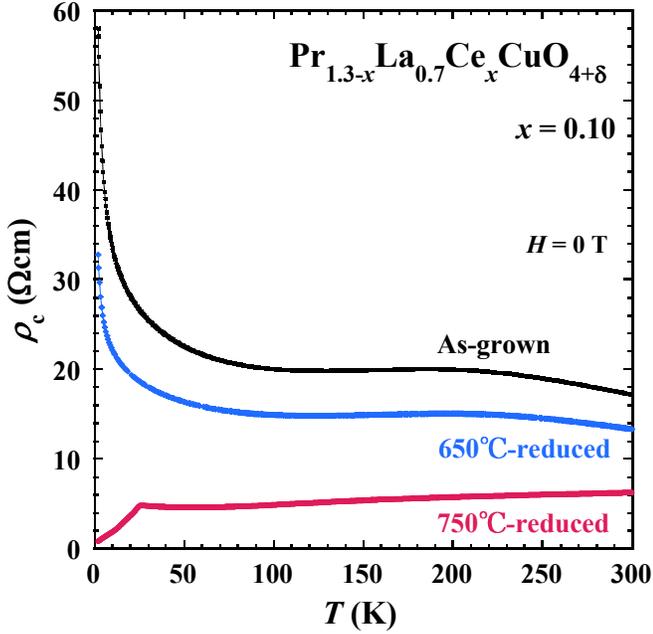}
\caption{(Color online) Temperature dependence of the c axis electrical resistivity $\rho_{\rm c}$ in zero field of Pr$_{1.3-x}$La$_{0.7}$Ce$_x$CuO$_{4+\delta}$ with $x=0.10$.  The data of the as-grown crystal and crystals reduced at 650$^{\rm o}$C and 750$^{\rm o}$C are shown.}
\label{f1}
\end{figure}

Following features were uncovered from the present results of PLCCO with $x=0.10$ and different $\delta$ values.
For the non-SC as-grown crystal, the pseudogap derived from the AF fluctuation is open below $\sim$ 200 K and carriers are strongly localized at low temperatures.
The negative magnetoresistance observed in the as-grown crystal might be explained taking into account the Zeeman effect in the strongly localized state.~\cite{fukuyama}
Through the reduction annealing, carriers are delocalized, the Kondo effect takes place at low temperatures and $T_{\rm c}$ reaches $\sim$ 20 K for the moderately reduced crystal, while the pseudogap disappears.

Here we propose novel electronic and magnetic states in the electron-doped T'-cuprates based on the strong electron correlation with the Mott-Hubbard gap, as schematically depicted in Fig. 3.
In the fully reduced ideal state without excess oxygen, the square-planer coordination of Cu and O in the T' structure gives rise to the decrease in the Madelung energy of the Cu$3d$ orbitals compared with the octahedral coordination in the T structure. 
As shown in Fig. 3(b), therefore, the energy of the upper Hubbard band (UHB) of the Cu$3d_{x^2-y^2}$ orbital becomes low enough to mix with the O$2p$ band.
That is, the charge-transfer gap between UHB of the Cu$3d_{x^2-y^2}$ and the O$2p$ band is collapsed, so that both hole and electron carriers simultaneously emerge at the Fermi level in the parent compounds of the T'-cuprates, carrying SC current.~\cite{naito}
Accordingly, it is speculated that the ideally reduced state of PLCCO with $x=0.10$ is a coexistent one of both an electron-doped state of UHB of Cu$3d_{x^2-y^2}$ and a hole-doped one of the O$2p$ band, as shown in Fig. 3(b).
In the CuO$_2$ plane shown in Fig. 3(a), Cu$3d_{x^2-y^2}$ spins of the lower Hubbard band (LHB) are antiferromagnetically correlated with each other. 
However, the AF long-range order is not formed because of mobile hole and electron carriers disturbing the AF ordering.

In the case of excess oxygen residing adjacent to the CuO$_2$ plane, two holes are produced in the CuO$_2$ plane by one excess oxygen.~\cite{onehole}
The doped holes in the CuO$_2$ plane tend to be localized near the excess oxygen due to the disorder of the electrostatic potential.
Moreover, the holes are doped into the Cu$3d$ and O$2p$ orbitals taking into account the strong electron correlation. 
This is because the energy of the Cu$3d$ orbitals is raised by the excess oxygen, as shown in Fig. 3(d).
The doped hole in the Cu$3d$ orbital resides not in LHB of Cu$3d_{x^2-y^2}$ but in UHB of Cu$3d_{3z^2-r^2}$ in order to gain the energy of the Zhang-Rice singlet of Cu$3d_{x^2-y^2}$ (LHB) and O$2p$ spins.
This is reasonable, because the splitting of Cu$3d_{x^2-y^2}$ and Cu$3d_{3z^2-r^2}$ in energy is smaller than the Mott-Hubbard gap of $\sim 8$ eV.
Accordingly, Cu$3d_{3z^2-r^2}$ (LHB) spins are probably induced at the Cu-site adjacent to the excess oxygen.
The induced Cu$3d_{3z^2-r^2}$ (LHB) spins probably behave as free spins, as schematically depicted in Fig. 3(c).
On the other hand, both hole and electron carriers tend to be localized because of the disorder of the electrostatic potential near the excess oxygen, resulting in the recovery of the AF order of Cu$3d_{x^2-y^2}$ (LHB) spins.
This is able to explain the increase in $\rho_{\rm ab}$ due to the strong localization and the hump of $\rho_{\rm c}$ due to the opening of the AF pseudogap.

As for the moderately reduced case, the localization of carriers is weak, so that itinerant carriers tend to be scattered by free Cu$3d_{3z^2-r^2}$ (LHB) spins, bringing about the Kondo effect.
Moreover, the AF order is destroyed by itinerant carriers, resulting in the disappearance of the AF pseudogap.

Here the consistency between the present model and formerly reported experimental results of the T'-cuprates is discussed.
The Kondo effect in the underdoped thin films of T'-cuprates observed by Sekitani {\it et al}.~\cite{sekitani} would appear due to the incomplete removal of excess oxygen. 
In the SC thin film of NCCO with $x=0.131$, log $T$-dependent $\rho_{\rm ab}$ has been observed at low temperatures due to the Kondo effect,~\cite{sekitani2} which is probably corresponding to the log $T$-dependent $\rho_{\rm ab}$ in the present 750$^{\rm o}$C-reduced PLCCO with $x=0.10$.
On the other hand, the pseudogap in NCCO with $x<0.14$ observed by Onose {\it et al}. would originate from the incomplete removal of excess oxygen, because the AF order is formed at low temperatures in their crystals showing the pseudogap.

The present model illustrated in Fig. 3 suggests the existence of dual carriers in the T'-cuprates.
In fact, Hall coefficient of PCCO thin films have revealed that dominant carriers change from electrons to holes through the Ce substitution, indicating the presence of dual carriers near the Fermi level.~\cite{dagan}
In order to justify the present model, measurements of the Hall coefficient would be necessary.

Finally, it is noted that simple pictures of the band metal without strong electron correlation~\cite{massidda,naito} are also able to explain the present results. 
In the fully reduced ideal state without excess oxygen, localized Cu spins are absent and itinerant electrons bring about superconductivity in the CuO$_2$ plane.
On the other hand, the introduction of excess oxygen gives rise to free Cu spins in the CuO$_2$ plane adjacent to the excess oxygen, leading to the Kondo effect. 
Moreover, a lot of excess oxygen makes electrons be localized, resulting in the generation of localized Cu spins and the formation of the AF order.
In order to clarify whether the T'-cuprates are band metals or strongly correlated electron systems, both muon-spin-relaxation measurements to investigate the Cu-spin state and angle-resolved-photoemission measurements to investigate the band structure are under contemplation.

\begin{figure}
\includegraphics[width=1.0\linewidth]{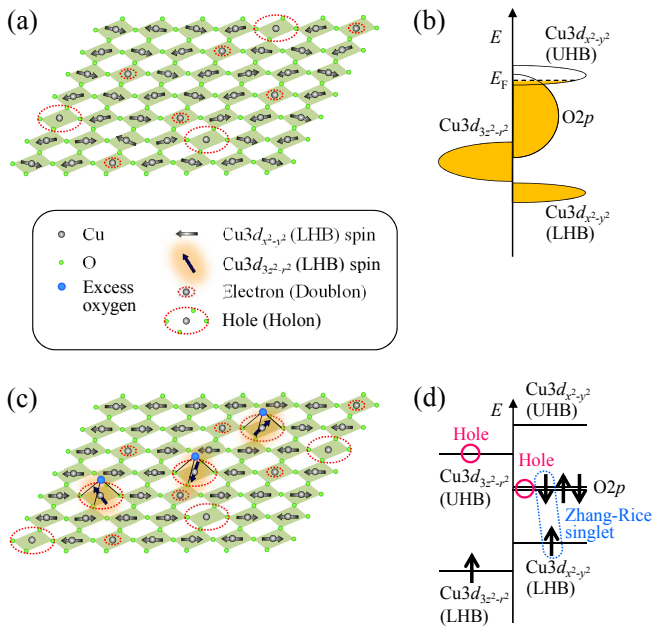}
\caption{(Color online) Schematic drawings of electron and hole carriers and Cu spins in the CuO$_2$ plane in (a) fully reduced and (c) as-grown Pr$_{1.3-x}$La$_{0.7}$Ce$_x$CuO$_{4+\delta}$ with $x=0.10$.  (b) Band structure of a fully reduced crystal corresponding to (a).  (d) Electronic structure of the CuO$_2$ plane near the excess oxygen.}
\label{f1}
\end{figure}

To be summarized, in electron-doped PLCCO single crystals with $x=0.10$, $\rho_{\rm ab}$ has revealed that two-dimensional variable-range-hopping conduction and negative magnetoresistance at low temperatures in the as-grown crystal changes to the log $T$ dependence and negative magnetoresistance in moderately reduced SC crystals through the reduction annealing. 
Moreover, it has been found that a hump of $\rho_{\rm c}$ at $\sim$ 200 K observed in the as-grown crystal disappears in moderately reduced SC crystals.
Based on the model taking into account the strong electron correlation, it has been concluded that, in the case that a lot of excess oxygen is included, the strongly localized state of carriers due to the disorder of the electrostatic potential leads to the formation of the AF order, while in the case that a small amount of excess oxygen is included, itinerant carriers destroy the AF order and the Kondo state due to free Cu$3d_{3z^2-r^2}$ (LHB) spins induced by excess oxygen is realized. 
In the T'-cuprates, the appearance of superconductivity through the reduction annealing is probably triggered by the destruction of the AF order owing to the increasing itinerancy of carriers through the removal of excess oxygen.
Accordingly, the complete removal of excess oxygen in the T'-cuprates would result in the appearance of superconductivity in a wide range of the Ce concentration including the parent compound of $x=0$.

\begin{acknowledgment}

Fruitful discussions with A. Fujimori, M. Ogata, H. Tsuchiura and H. Yokoyama are gratefully acknowledged. 
We would like to thank M. Fujita for giving us useful information on the PLCCO crystal growth.
We are indebted to M. Ishikuro of the Institute for Materials Research (IMR), Tohoku University, Japan, for his help in the ICP analysis. 
We are also grateful to A. Haneda of the Technical Division, School of Engineering, Tohoku University, Japan, for her aid in the SEM measurements.
The high magnetic field experiments were supported by the High Field Laboratory for Superconducting Materials (HFLSM), Institute for Materials Research, Tohoku University, Japan.
This work was carried out at 'Exploratory Research Program for Young Scientists in the Graduate School of Engineering, Tohoku University', Japan. 

\end{acknowledgment}


\begin{thebibliography}{99}
\bibitem{onehole}
Depending on the bonding state of the excess oxygen, one hole may be produced by one excess oxygen in the CuO$_2$ plane.  In this case, the doped hole resides in UHB of Cu$3d_{3z^2-r^2}$ or the O$2p$ band, giving rise to Cu$3d_{3z^2-r^2}$ free spins and bringing about the Kondo effect.
\end{thebibliography}
\end{document}